# Feedback-based active reset of a spin qubit in silicon

T. Kobayashi[1,2, a)], T. Nakajima[2], K. Takeda[2], A. Noiri[2], J. Yoneda[2,b)], and S. Tarucha[1,2]

[1]*RIKEN Center for Quantum Computing, Wako, Saitama 351-0198, Japan*

[2]*Center for Emerging Matter Science, RIKEN, Wako, Saitama 351-0198, Japan*

Feedback control of qubits is a highly demanded technique for advanced quantum information protocols such as quantum error correction. Here we demonstrate active reset of a silicon spin qubit using feedback control. The active reset is based on quantum non-demolition readout of the qubit and feedback according to the readout results, which is enabled by hardware data processing and sequencing. We incorporate a cumulative readout technique to the active reset protocol, enhancing initialization fidelity above a limitation imposed by accuracy of the single QND measurement fidelity. Based on an analysis of the reset protocol, we suggest a way to achieve the initialization fidelity sufficient for the fault-tolerant quantum computation.

---

[a)] takashi.kobayashi@riken.jp

[b)] Current address: Tokyo Tech Academy for Super Smart Society, Tokyo Institute of Technology, Meguro-ku, Tokyo 152-8552, Japan.

Silicon-based spin qubits are regarded as a powerful candidate for the building blocks of scalable quantum information processors owing to the high quantum-gate fidelities [1–6], high-temperature compatibility [7,8], and well-developed fabrication technologies [9,10]. Fundamental technologies for a small number of qubits have matured as represented by the demonstration of a quantum error correction [6], and technologies for scaling up toward quantum information processors are more focused recently [11]. Feedback control where qubits are controlled conditionally on results of quantum non-demolition (QND) measurements is one of the key technologies for scaling up, employed in a proposal of the surface code [12]. Active reset of a qubit using is the most fundamental feedback control, demonstrated in varieties of qubit platforms [13–15] and, very recently, in a silicon quantum-dot qubit system [11]. The initialization fidelity of active reset depends on a protocol to generate feedback. A good protocol will facilitate increasing initialization fidelity over the target value > 99.5 % for a fault-tolerant quantum computer [16].

In this work, we report feedback-based active reset of an electron spin qubit in a silicon quantum dot. The spin-qubit state is read out by QND measurements [17–19], whose outcome serves to generate feedback to reset the qubit. A combination of a digital signal processing (DSP) hardware [20] and a hardware sequencer [21] enables us to generate feedback much shorter than spin relaxation time $T_1$. First, we have tested feedback generation using a simple reset protocol based on a single QND measurement and have confirmed proper operations of the protocol. To improve the initialization fidelity, we focus on the measurement to generate feedback, whose fidelity is critical to generate feedback appropriately. We incorporate a cumulative readout technique to the feedback generation logic [17,18] and obtain the initialization fidelity of 98.3 %. We have analyzed the reset protocol and propose a pathway to achieve the initialization fidelity higher than 99.5 %.

**Results**

**Active reset protocol.** Figure 1a outlines concept of an active reset protocol to initialize a qubit to the ground (spin-down) state. The protocol requires an auxiliary qubit (ancilla qubit) in addition to the qubit to be initialized (data qubit). The initial two-qubit state is represented as $|0_A\rangle|\psi_D\rangle$ where $|s_X\rangle$ denotes the spin-down ($s_X = 0_X$) and -up ($s_X = 1_X$) states of the ancilla (X = A) and data (X = D) qubits and $|\psi_D\rangle$ represents an arbitrary superposition state $\alpha|0_D\rangle + \beta|1_D\rangle$ with $|\alpha|^2 + |\beta|^2 = 1$. We first apply a QND measurement (blue area) consisting of a controlled rotation (CROT) gate using the data qubit as the controlled bit and a subsequent destructive measurement to the ancilla qubit. The CROT gate entangles the two qubits to, for example, $\alpha|1_A\rangle|0_D\rangle + \beta|0_A\rangle|1_D\rangle$, ignoring change in relative phase not affecting the final result. The two-qubit state is first entangled to, for example, $\alpha|1_A\rangle|0_D\rangle + \beta|0_A\rangle|1_D\rangle$ by the CROT gate ignoring relative-phase change irrelevant to the result. The subsequent destructive measurement projects the entangled state to $|0_A\rangle|1_D\rangle$ or $|1_A\rangle|0_D\rangle$ with probabilities $|\alpha|^2$ and $|\beta|^2$, and yields an ancilla measurement outcome $\mu = 0_A$ or $1_A$ without disturbing the data-qubit state. Because of the correlation between the $|0_A\rangle$ and $|1_D\rangle$ ($|1_A\rangle$ and $|0_D\rangle$) states after the projection, we obtain an estimator $m = 1_D$ ($0_D$) of the data qubit state from the outcome $\mu = 0_A$ ($1_A$) in a QND manner. The estimator $m$ is fed back to the data-qubit state through the following π-rotation gate, which is classically conditioned so as to π-rotate the data qubit only when $m = 1_D$ is obtained (yellow area). After this rotation, the data qubit is reset to the $|0_D\rangle$ state and arbitrary quantum operations can be performed subsequently.

This reset protocol relies on the QND readout and the conditioned π rotation. The initialization fidelity relates to other fidelities as $F_I = 1/2 + (2F_{fb,R} - 1)F_G(2F_{QND} - 1)/2$ (See Methods). Here, $F_I$ is the initialization fidelity, $F_{fb,R}$ is the fidelity of the QND readout of the data qubit to generate feedback, $F_{QND}$ is the state-preservation fidelity for the data qubit during the reset protocol (QND fidelity) [22], and $F_G$ is the π-rotation fidelity. $F_I$, $F_{fb,R}$, and $F_{QND}$ are averaged over input data-qubit states. $F_{fb,R}$ is degraded not only by errors in ancilla destructive measurement but also by CROT-gate errors to map the data-qubit state on the ancilla-qubit state. In silicon spin qubits, $F_{fb,R}$ is typically much lower than $F_G$ and $F_{QND}$; $F_{fb,R} \leq 92$ % and $F_{QND} \geq 99$ % for the reset protocol based on a single QND measurement as shown later. In this regime, the relation between $F_I$ and other fidelities is approximated by $F_I \approx F_{fb,R}$ and thus $F_I$ is expected to increase with $F_{fb,R}$ in the active reset protocol.

An important feature of the QND readout is preservation of the measured qubit state, which enables us to improve $F_{fb,R}$ using cumulative readout techniques [17–19]. To incorporate the cumulative readout to the active reset protocol, we design a quantum circuit (Fig. 1b) which can replace the single QND measurement in Fig. 1a. A repetition of the measurements provides a set of ancilla measurement outcomes $\{\mu\}_N$ ($N$ is number of repetition). A cumulative estimator $M_N$ of the data-qubit state is obtained from the set and used to condition the subsequent π-rotation gate. We note that the cumulative estimator $M_N$ must be calculated in real time to generate feedback in contrast with the previous demonstrations of cumulative readout [17–19]. This requires a more sophisticated feedback-generation system than active reset based on a single QND measurement [11,13].

**Qubit system and feedback control.** The ancilla and data qubits are hosted by the left and right quantum dots in a double quantum dot device [18,23] (Fig. 1c, dashed box) with an adjacent charge sensor to read charge occupations (See Methods). An external magnetic field splits qubit levels by roughly 16 GHz (Figs. 2 and 4) or 19 GHz (Fig. 3). A micromagnet on top of the device induces a slanting magnetic field to couple the spin qubits to microwave (MW) oscillating voltage applied to a gate (barrier gate), which drives the ancilla and data qubits at Rabi frequencies of 2-4 MHz. The micromagnet also induces a difference in the Zeeman splittings of the ancilla and data qubits of around 600 MHz large enough to address them individually. The exchange interaction between the qubits is turned on and off by a fast barrier-gate bias (See Supplementary Information for details of the experimental time sequence). As the exchange coupling in the on state (6-9 MHz) is much smaller than the Zeeman splitting difference, it is effectively an Ising-type interaction [18] which splits the $|0_A\rangle|0_D\rangle \leftrightarrow |1_A\rangle|0_D\rangle$ and $|0_A\rangle|1_D\rangle \leftrightarrow |1_A\rangle|1_D\rangle$ transition energies. A π rotation using a MW pulse resonant to one of these transitions works as a CROT gate. Here, the choice of the data-qubit projection axis in the QND measurement enables us to ignore conditional phase factors accumulated by pulsing barrier gate bias and off-resonance drive of the ancilla qubit [3]. $T_1$ of the data qubit is > 6 ms in this work. The sample is cooled down at a dilution refrigerator with a base temperature of 26 mK. The electron temperature is 36 mK.

We test conditioning of the π-rotation gate to flip back the data-qubit state to $|0_D\rangle$ using a feedback scheme based on an estimator of a single QND measurement *m* in the actual experimental setup (Fig. 1c). The ancilla destructive readout is performed by a combination of spin-selective tunneling and reflectometry charge sensing (Elzerman readout [24]). Figure 1d shows typical time-domain signals in the measurement span between 30 and 130 μs after the CROT gate. Presence of a dip (arrows) in this time span is an indication of the $|1_A\rangle$ state and thus $\mu = 1_A$. Accordingly, the DSP hardware assigns the time-domain signals with and without the dip to QND measurement estimators $m = 0_D$ or $1_D$, respectively, in the data-processing time span (between 130 μs and 205 μs, not shown). The obtained estimators are used to condition the π-rotation gate by an arbitrary waveform generator (AWG) whose output is supplied to a MW switch placed on the output port of a MW generator. Figure 1e shows AWG output signals following each of the readout signals shown in the same color in Fig. 1d. The voltage level in the time span reserved for the π pulse (between 210 and 220 μs) reflects the presence of the dip in each readout signal; only when the dip is absent, that is, $m_{fb} = 1_D$ is obtained, the AWG sets the voltage level high to turn on the MW switch to transmit the π-rotation pulse. These output signals show that the feedback signals are appropriately generated according to the readout signal. We note that we use shorter spans for readout, data processing, and the π pulse in the following.

**Implementation of the active reset using single QND readout.** We now demonstrate the qubit reset protocol using the QND measurement and the feedback scheme. Figure 2a shows the quantum circuit used. Before the reset protocol, the ancilla and data qubits are prepared in the $|0_D\rangle$ and $|0_A\rangle$ states by reloading an electron with down spin to each quantum dot. The data qubit is subsequently subjected to a π/2 pulse, which makes a superposition state of the $|0_D\rangle$ and $|1_D\rangle$ states. The following QND measurement at the beginning of the reset protocol randomly projects the data-qubit state to $|0_D\rangle$ or $|1_D\rangle$, which works as an unbiased input state to evaluate the reset protocol. To analyze fidelities, we apply a resonant MW burst to the data qubit for duration $\tau_b$ after the conditional $\pi$ rotation and read it out by another QND measurement (outcome $\mu$) and a destructive measurement to the data qubit (outcome $m_d$). Here time taken by the reset protocol $\tau_{\text{reset}}$ is 93.5 μs, measured from the CROT gate to the conditional $\pi$ rotation.

Figure 2b shows Rabi oscillations of the data qubit exhibited by $m_d$ (blue) and a QND estimator $m$ obtained from the outcome $\mu$ (orange). For comparison, we also perform measurements without the feedback (open circles), using the same pulse sequence but suppressing conditional $\pi$-rotation pulse intentionally. The data set with feedback shows clear Rabi oscillations with visibility of 44% and 41 % for $m_d$ and $m$ (from fitting, solid curves). We note that weak bias in the data-qubit state before the reset protocol, which is manifested as the finite but low visibility of the Rabi oscillations in $m_d$ without the feedback (2.8 %, dashed curve), cannot account for the visibility with feedback. This result indicates successful initialization of the data qubit by the reset protocol.

We analyze joint probabilities for $m$ and $m_d$ to extract the initialization fidelities of the active reset protocol. Figure 2c shows joint probabilities $P(m,m_d)$ for all four possible combinations of $m$ and $m_d$ (circles): $(m, m_d) = (0_D, 0_D), (0_D, 1_D), (1_D, 0_D),$ and $(1_D, 1_D)$. $P(m,m_d)$ is expressed in a form where the infidelities in preparation and measurement of the qubit state are separated [17,18]:

$$P(m, m_d) = \{1 - p(\tau_b)\}\Theta_{0,m}(f_{R,0})\Theta_{0,m_d}(f_{d,R,0}) + p(\tau_b)\Theta_{1,m}(f_{R,1})\Theta_{1,m_d}(f_{d,R,1}).$$

Here, $f_{R,s}$ ($f_{d,R,s}$) denotes the readout fidelities of the QND measurement (the destructive measurement) for the data qubit state prepared in $|0_D\rangle$ ($s = 0$) or $|1_D\rangle$ ($s = 1$) after the resonant MW burst. $\Theta_{0,m}(x) = x$ for $m = 0_D$ and $\Theta_{0,m}(x) = 1 - x$ for $m = 1_D$, and always $\Theta_{1,m}(x) = 1 - \Theta_{0,m}(x)$. $p(\tau_b)$ is the $|1_D\rangle$-state probability distribution after the resonant MW burst for $\tau_b$, modeled as $p(\tau_b) = B - A\cos(2\pi f_{Rabi}\tau_b)\exp(-\tau_b/T_{2,Rabi})$ using Rabi frequency $f_{Rabi}$, Rabi-oscillation decay time $T_{2,Rabi}$, amplitude $A$, and offset $B$. Fitting $P(m,m_d)$ to the expression using a maximum likelihood method (curves), we obtain $A = 0.31$, $B = 0.51$, $f_{R,0} = 0.85$, $f_{R,1} = 0.83$, $f_{d,R,0} = 0.95$, $f_{d,R,1} = 0.71$. The amplitude $A$ relates solely to the initialization fidelity $F_I$ as $F_I = A + 1/2$, yielding $F_I = 81$ %. While we do not have direct evaluation of the $F_{fb,R}$, the QND readout at the end is performed in the manner same as it. Thus the averaged fidelity $f_R = (f_{R,0} + f_{R,1})/2 = 84$ % should coincide with $F_{fb,R}$. The $F_I$ value close to $F_{fb,R}$ implies that $F_I$ is limited by $F_{fb,R}$ as expected from their relation.

The destructive measurement of the data qubit used in Fig. 2a requires an electron reservoir [24], which imposes geometrical constraints in quantum information architectures. To avoid usage of an electron reservoir, we implement a quantum circuit equivalent to Fig. 2a but without the destructive measurement by repeating a cycle consisting of a QND measurement, a conditional π-rotation gate, and a resonant MW burst to the data qubit for $\tau_b$ (Fig. 3a). $\tau_{reset}$ is 93 µs out of 100 µs taken by each cycle. Every ancilla measurement provides an outcome µ and yields an estimator for the data-qubit state $m$. The estimator is used to generate feedback for the next cycle in real time and also to calculate probability distributions for the data-qubit state at the end of each cycle, for which the data-qubit destructive readout is used in Fig. 2a. Figure 3b shows Rabi oscillations exhibited by $m$ similarly to Fig. 2b but measured by using the quantum circuit in Fig. 3a. Clear difference between presence (blue) and absence (gray) of the feedback indicates successful reset of the data qubit by this quantum circuit. We note that, while the data qubit is free from electron reload throughout the experiment in contrast to Fig. 2a, the ancilla qubit still uses electron reload for a destructive measurement. The electron reload can be avoided completely by using Pauli spin blockade readout as demonstrated in ref. [11] although an additional ancilla spin is required for a QND measurement.

**Feedback using cumulative readout.** We attempt to improve $F_I$ by incorporating a cumulative readout technique [17–19] to the active reset protocol. Increase of the readout fidelity by cumulative readout is tested using a quantum circuit shown in Fig. 3c. 20-times QND measurements are performed between the resonant MW burst and the QND measurement for the reset protocol in comparison with Fig. 3a, resulting in a set of ancilla readout outcomes $\{\mu\}_{20}$. Similarly to Fig. 2c, fidelities can be extracted by analyzing joint probabilities $P(m_k, m_{20})$ where $m_k$ ($k = 1\text{-}20$) is the QND estimator for the data-qubit state obtained from the single outcome $\mu_k$. We obtain fidelities of the individual QND measurement for the $|0_D\rangle$ and $|1_D\rangle$ states, $f_{R,0} = 93\%$, $f_{R,1} = 89\%$, their average $f_R = 91\%$, and the initialization fidelity $F_I = 88\%$ (See Supplementary Information for the joint probability analyses). Also, from the $k$ dependence of $P(m_k, m_{20})$, we can also extract $T_1$'s for the $|0_D\rangle$ and $|1_D\rangle$ states $T_{1,0} = 80\pm20$ ms and $T_{1,1} = 6.6\pm0.1$ ms. Using a Bayesian estimation method taking these $T_1$'s into account, we calculate cumulative estimators $M_n$ for the data-qubit state before the measurement from subsets $\{\mu\}_n = \{\mu_1, \mu_2, ..., \mu_n\}$ ($n \leq 20$) of the $\{\mu\}_{20}$ (See Supplementary Information for the analyses of the repetitive measurement outcomes). Figure 3d shows the Rabi oscillations obtained from the estimators $M_n$ for $n = 1$ and 20 (blue and orange). The Rabi-oscillation visibility for $n = 20$ is higher than $n = 1$, implying improved readout fidelities by cumulative readout. Given $F_I = 88\%$, we can estimate the cumulative readout fidelities for the $|0_D\rangle$ and $|1_D\rangle$ states $F_{R,0}$, and $F_{R,1}$, and their average $F_R$ as a function of $n$ (Fig. 3e). $F_R$ increases with $n$ for $n < 10$ and saturates to 97% for higher $n$. The cumulative readout certainly yields a higher fidelity than $f_R$ for the single QND measurement.

Figure 4a shows an implementation of the feedback-based reset based on the cumulative readout. The single QND readout to generate feedback in Fig. 3d is replaced by a 11-fold repetition of the QND measurements. Each of the QND readout signals is transformed to ancilla readout outcome $\mu_{fb,n}$ ($n = 1\text{-}11$) by the DSP hardware immediately after each QND measurement and transferred to a Bayesian-estimation block constructed in the hardware sequencer (the box denoted by B, also see Supplementary

Information for details of this logic block). The Bayesian-estimation block updates the cumulative estimator of the data-qubit state $M_{fb,n}$ online (this is in contrast with the offline Bayesian estimation used in Fig. 3d-f). The feedback to condition the π rotation is generated after the 11th QND measurement according to $M_{fb,11}$. The Bayesian-estimation block takes likelihood parameters into account but not $T_1$'s due to register constraints, meaning that estimation errors due to spin relaxation during the repetitive measurements cannot be amended. While the individual cycle of the repetition takes shorter (65 μs) than Fig. 3c, $\tau_{\text{reset}}$ is increased to 708 μs due to the repetitive measurements. We also perform another 20-fold repetition of the QND measurements and acquire $\{\mu\}_{20}$ for analysis.

Fidelities is evaluated by the joint probability analysis using $\{\mu\}_{20}$ in the same manner as Fig. 3c-e; we obtain $f_{R,0} = 94\pm1$ %, $f_{R,1} = 90.0\pm0.7$ %, $f_R = 91.7\pm0.8$ %, $F_I = 98.33\pm0.08$ %, $T_{1,0} = 130\pm30$ ms, and $T_{1,1} = 19.8\pm0.6$ ms. The initialization error rate (1.67±0.08 %) is five times lower than the averaged readout error rate (8.3±0.8 %), which approximately coincides with the initialization error rate for the reset protocol based on the single QND measurement. The cumulative estimator $M_{20}$ from $\{\mu\}_{20}$ calculated by a Bayesian method taking $T_{1,0} = 130$ ms and $T_{1,1} = 19.8$ ms into account exhibits Rabi oscillations (Fig. 4b) more pronouncedly than Fig. 3d. Incorporating cumulative readout to generate feedback, we successfully obtain a $F_I$ value beyond readout fidelities of the individual QND readout, which limits $F_I$ in the protocol based on the single QND measurement.

To address the residual initialization error, we review the cumulative readout process more deeply. Analysis of the joint probability $P(m_k, m_{20})$ also provides the fidelities of state preservation after a $k$-fold repetition of the QND measurements for the $|0_D\rangle$ and $|1_D\rangle$ states, $F_{\text{QND},0}$ and $F_{\text{QND},1}$, and their average $F_{\text{QND}}$ (Fig. 4c). While the observed $F_{\text{QND}}$ is higher than the previous report [18], it is comparable with $F_I$ at $k = 11$ ($F_{\text{QND}} = 98.2\pm0.4$ %). The cumulative readout fidelities $F_{R,0}$, $F_{R,1}$, and $F_R$ can be estimated from outcome subsets $\{\mu\}_n$ by the Bayesian method same as Fig. 4b, providing $F_R = 98.7\pm0.8$ % at $n = 11$ which should coincide with $F_{fb,R}$ (Fig. 4d). Since the saturation of fidelities at large $n$ is attributed to state-preservation errors, higher $F_{QND,0}$ and $F_{QND,1}$ should result in higher $F_R$. Since the Bayesian logic used to generate feedback does not include spin relaxation, we also inspect the cumulative readout fidelities estimated by Bayesian method assuming infinitely long $T_1$'s (Fig. 4e). $F_{R,0}$ and $F_{R,1}$ deviate from their estimations with the measured $T_1$'s at large $n$ due to spin relaxation (Fig. 4e inset). Nevertheless, $F_R$ is almost unaffected by the difference in the estimation method (98.6±0.8 %). Reviewing the cumulative readout process, the initialization fidelity is likely to increase with the QND fidelities.

**Discussion**

The time required to reset a qubit is important for improving the throughput of quantum information processing. In this work, we can decrease $\tau_{reset}$ for the reset protocol using the single QND measurement (Fig. 3a) to ≈60 μs, which is mainly limited by the destructive readout taking 40 μs. The Elzerman readout is generally limited in terms of the readout speed as it relies on stochastic electron exchange between a quantum dot and an electron reservoir. By employing the spin readout based on the Pauli spin blockade, readout time can be decreased to 1 μs with high fidelity [25–28]. The second dominant limitation is the time to process a data set and to set the AWG output. As this takes slightly longer than 10 μs in the present setup, we reserve 15 μs for secure feedback generation (See Supplementary Information for details of the experimental time sequence). We suppose this latency is due to the sequencing hardware and expect that it can be reduced to less than 1 μs as demonstrated in ref. [11]. These potential improvements will decrease $\tau_{reset}$ for feedback based on a single QND measurement to 2 μs or less.

Incorporation of the cumulative readout to the feedback-based reset is a reasonable way to enhance $F_I$ exceeding the fault-tolerant threshold for initialization 99.5 % [16], because requirement for the readout fidelity of a single QND measurement is not as high as the fault tolerant threshold. While the time-consuming repetition of QND measurements is a drawback of this scheme, a numerical simulation predicts that, three-fold repetition provides $F_{fb,R} > 99.9$ % if the readout fidelity of the individual QND measurement is 99 % [17]. Together with the short readout time (1 μs) and the short data-processing time (<

1 μs), this enables to perform high fidelity cumulative readout with $\tau_{\text{reset}}$ of 6 μs. Such a short $\tau_{\text{reset}}$ yields $F_{\text{QND}} > 99.98$ % assuming the $T_1$'s observed in Fig. 4. The $F_{\text{I}}$ of the active reset protocol based on the cumulative readout for these envisioned $F_{fb,\text{R}}$ and $F_{\text{QND}}$ values is calculated as $F_{\text{I}} > 0.5 + 0.499 F_{\text{G}}$ and thus $F_{\text{I}} > 99.5$ % for $F_{\text{G}} > 99.2$ %. As the single-qubit gate fidelity of 99.5 % is routinely obtained in state-of-the-art spin qubits in silicon [1–6,11,23], $F_{\text{I}}$ higher than 99.5 % is achievable by improving each QND readout fidelity and shortening the reset protocol.

In conclusion, we have demonstrated a deterministic initialization scheme of a spin qubit based on the QND measurement. Fast data processing and sequencing enable us to implement feedback according to the QND-readout estimators before the data-qubit state has changed due to spin relaxation. This scheme works properly regardless of isolation of a qubit from the electron reservoirs. We also find that cumulative measurement techniques can be incorporated to improve the initialization fidelity by the reset protocol. This scheme opens a pathway to develop silicon spin quantum information architectures suitable for scaling up demanded for quantum information processors.

**Methods**

**Initialization fidelity of the active reset protocol.** The averaged initialization fidelity $F_{\text{I}}$ of the active reset protocol to the $|0_{\text{D}}\rangle$ state is expressed as follows:

$$F_I = (F_{I,0} + F_{I,1})/2,$$

$$1 - F_{I,0} = (1 - F_{fb,R,0})F_G F_{QND,0} + (1 - F_{fb,R,0})(1 - F_G)(1 - F_{QND,0}) + F_{fb,R,0}(1 - F_{QND,0}),$$

$$1 - F_{I,1} = (1 - F_{fb,R,1})F_{QND,1} + F_{fb,R,1}(1 - F_G)F_{QND,1} + F_{fb,R,1}F_G(1 - F_{QND,1}).$$

Here, $F_{\text{I},s}$, $F_{fb,\text{R},s}$, and $F_{\text{QND},s}$ are the initialization, readout and QND fidelities for a given data-qubit state $|s_{\text{D}}\rangle$ ($s_{\text{D}} = 0_{\text{D}}, 1_{\text{D}}$), respectively, and $F_{\text{G}}$ is the fidelity of the conditional π rotation applied to the data qubit. Substituting the second and third equations to the first one, we obtain

$$F_I = \frac{1 + F_{QND,0} - F_{QND,1} - F_G(2F_{QND,0} - 1)}{2} + \frac{1}{2}F_{fb,R,0}F_G(2F_{QND,0} - 1) + \frac{1}{2}F_{fb,R,1}F_G(2F_{QND,1} - 1).$$

Using $F_{fb,R} = (F_{fb,R,0}+F_{fb,R,1})/2$ and $F_{QND} = (F_{QND,0} + F_{QND,1})/2$,

$$F_I = \frac{1}{2} + \frac{1}{2}(2F_{fb,R} - 1)F_G(2F_{QND} - 1) + \frac{1}{2}\{(F_{fb,R,0} - F_{fb,R,1})F_G + (1 - F_G)\}(F_{QND,0} - F_{QND,1}).$$

Since $F_{QND,0} - F_{QND,1} \ll 1$ (1.4 % for Fig. 3 and 2.8 % for the cumulative readout in Fig. 4) in our experiments, we neglect the third term and obtain

$$F_I = \frac{1}{2} + \frac{1}{2}(2F_{fb,R} - 1)F_G(2F_{QND} - 1).$$

In the above argument, random data-qubit states are input to the reset protocol. This is not the case in the experiments presented in Figs. 3 and 4, since the input data-qubit state may be strongly correlated to the final data-qubit state of the previous cycle. If general data-qubit states are input, initialization fidelity is represented by using probability distribution of the $|1_D\rangle$ state before the reset protocol, $p_{0,1}$:

$$F_I = (1 - p_{0,1})F_{I,0} + p_{0,1}F_{I,1} = F_{I,0} + p_{0,1}\delta F_I,$$

using $\delta F_I = F_{I,1} - F_{I,0}$. If the initialization fidelity is state dependent ($\delta F_I \neq 0$), the second term must be considered in estimation of $F_I$. Probability distribution of the $|1_D\rangle$ state, $p_1(\tau)$, is represented by $F_{I,s}$ as

$$p_1(\tau) = F_I \rho(\tau) + (1 - F_I)(1 - \rho(\tau)) = \left(F_I - \frac{1}{2}\right)(2\rho(\tau) - 1) + \frac{1}{2}.$$

$$\rho(\tau) = \frac{1}{2} - \frac{1}{2}\cos 2\pi f_{Rabi}\tau \, e^{-\tau/T_{2,Rabi}}.$$

In the present experiment, the data qubit is not subjected to spin reload between each Rabi burst and reset protocol. In this case, the $p_{0,1}$ is approximated by $p_1(\tau)$, and $p_1(\tau)$ is represented by the initialization fidelities as

$$p_1(\tau) = \frac{\left(F_{I,0} - \frac{1}{2}\right)(2\rho(\tau) - 1) + \frac{1}{2}}{1 - \delta F_I(2\rho(\tau) - 1)}.$$

Assuming small data-qubit-state dependence of the initialization fidelity, that is, $\delta F_I \ll 1$, we can approximate $\frac{1}{1-\delta F_I(2\rho(\tau)-1)} \approx 1 + \delta F_I(2\rho(\tau) - 1)$ and obtain

$$p_1(\tau) \approx \frac{1}{2} + \left(F_I - \frac{1}{2}\right)(2\rho(\tau) - 1) + \left(F_{I,0} - \frac{1}{2}\right)\delta F_I(2\rho(\tau) - 1)^2.$$

More precisely, we need to consider relaxation of the data qubit during the measurement after Rabi drive, which is not negligible for the repetitive measurement used in the experiments in Figs. 3c-e and 4. In this case, $p_{0,1}$ relates to $p_1(\tau)$ as $p_{0,1} = F_{QND,1} p_1(\tau) + (1 - F_{QND,0})(1 - p_1(\tau))$ and results in

$$p_1(\tau) \approx \frac{1}{2} + \left(F_I - \frac{1}{2} + \frac{\delta F_I (F_{QND,1} - F_{QND,0})}{2}\right)(2\rho(\tau) - 1)$$

$$+ \left(F_{I,0} - \frac{1}{2} + (1 - F_{QND,0})\delta F_I\right) \delta F_I (2F_{QND} - 1)(2\rho(\tau) - 1)^2.$$

The $|1_D\rangle$ probability distribution shown in Figs. 3b,d, and 4b should behaves like $p_1(\tau_b)$. However we do not observe oscillations originating from the third term, that is, doubled-frequency oscillations due to $(2\rho(\tau_b) - 1)^2$ (See Supplementary Information for the analyses of the repetitive measurement outcomes). This indicates that $\delta F_I$ is negligible in the experiments, and thus the $F_I$ in Figs. 3 and 4 can be estimated without considering probability distribution before the reset protocol.

**Experimental parameters.**

The device is a double quantum dot fabricated on a silicon/silicon-germanium heterostructure with the natural isotope abundance, which was investigated in previous reports [18,23]. Charge occupations of the double dot are read by a radio-frequency reflectometry technique with the charge sensor neighboring the left quantum dot [25,26]. The carrier frequency is 205 MHz and the carrier power at the output port of a radio-frequency signal generator is 13 dBm, which is attenuated to ≈−100 dBm before the device by an attenuator chain. The external magnetic field is 0.49 T (Figs. 2 and 4) or 0.60 T (Fig. 3). Zeeman splitting is 15.448 GHz (Fig. 2), 18.581 GHz (Fig. 3), and 15.438 GHz (Fig. 4) for the ancilla qubit when the exchange interaction is turned on, and is 16.006 GHz (Fig. 2), 19.156 GHz (Fig. 3), and 16.033 GHz (Fig. 4) for the data qubit when the exchange interaction is turned off. The difference in the Zeeman splittings of the ancilla and data qubits is around 600 MHz. The exchange coupling in the on state is 9.0 MHz (Fig. 2), 8.8 MHz (Fig. 3), and 6.1 MHz (Fig. 4). The Rabi frequency of the ancilla qubit is 2.8 MHz (Fig. 2), 2.2 MHz (Fig. 3), and 2.0 MHz (Fig. 4). The $|0_A\rangle|0_D\rangle \leftrightarrow |1_A\rangle|0_D\rangle$ and $|0_A\rangle|1_D\rangle \leftrightarrow |1_A\rangle|1_D\rangle$ transitions are used for the CROT gate in Figs. 2,4 and Fig. 3, respectively. The sizes of the exchange interaction and the ancilla Rabi frequency is not exactly tuned to cancel off resonance drive in the CROT gate, which slightly decreases $F_R$ of an individual QND measurement. In the experiments, the device is operated near the charge-symmetry point of the double dot to decouple the qubits from noise in energy level detuning. Dephasing time $T_2^*$ is approximately 1 μs for both qubits.

**Error analysis**

All uncertainties represent 1σ confidence intervals obtained from fitting.

**Acknowledgements**

This work was supported financially by Core Research for Evolutional Science and Technology (CREST), Japan Science and Technology Agency (JST) (JPMJCR15N2 and JPMJCR1675), PRESTO grant Nos. JPMJPR2017 and JPMJPR21BA, MEXT Quantum Leap Flagship Program (MEXT Q-LEAP) grant No. JPMXS0118069228, JST Moonshot R&D grant no. JPMJMS2065, and JSPS KAKENHI grant Nos. 17K14078, 18H01819, 19K14640, and 20H00237.

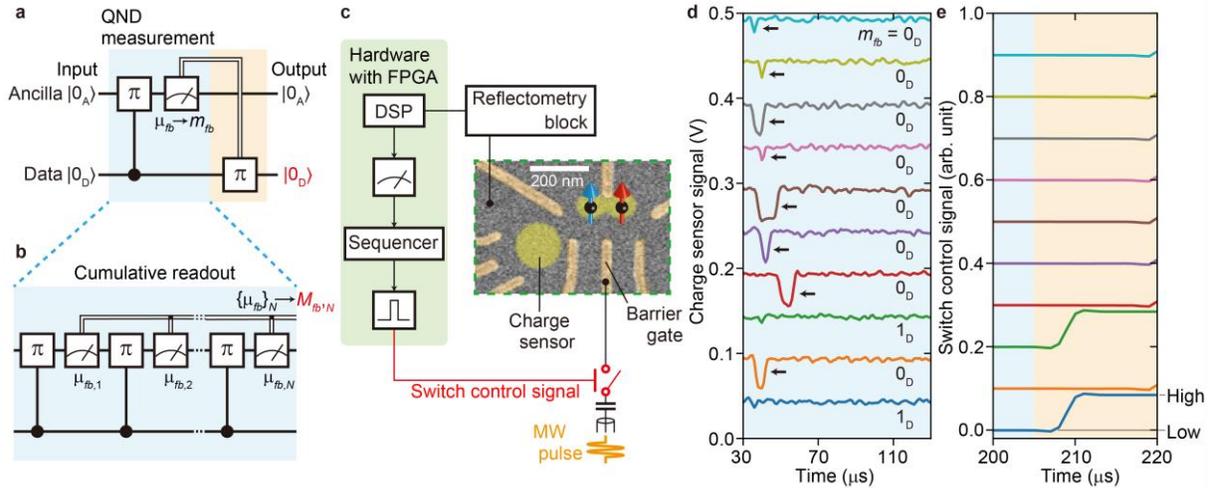

**Fig. 1 Implementation of the feedback-based active reset of a spin qubit in silicon**. **a,** Quantum circuit showing the basic reset protocol. The QND measurement of the data qubit is implemented by the CROT gate and a destructive measurement of the ancilla qubit (blue area). The data-qubit state is estimated at $m_{fb}$ according to the measurement outcome $\mu_{fb}$, a π-rotation gate (yellow area) is conditioned so as to flip the data qubit only when the estimator $m_{fb} = 1_D$. **a,** Quantum circuit for cumulative readout. QND measurements are repeated $N$ times, resulting in a set of outcomes $\{\mu_{fb}\}_N = \{\mu_{fb,1}, \mu_{fb,2}, ..., \mu_{fb,N}\}$. A cumulative estimator for the data qubit is obtained from $\{\mu_{fb}\}_N$. **c,** Schematic diagram of the experiment setup. A SEM image of the two-qubit device is shown in the green dashed box. **d,e,** Time-domain charge sensor signals of the destructive readout (d), and switch control signals output after the corresponding charge sensor signals in d (e). Black horizontal arrows indicate dips in the charge sensor signals due to reloading of the ancilla electron from the electron reservoir. These dips should appear when the ancilla qubit is in the $|1_A\rangle$ state, indicating that the data qubit is likely in the $|0_D\rangle$ state. The detected data qubit state is denoted in d.

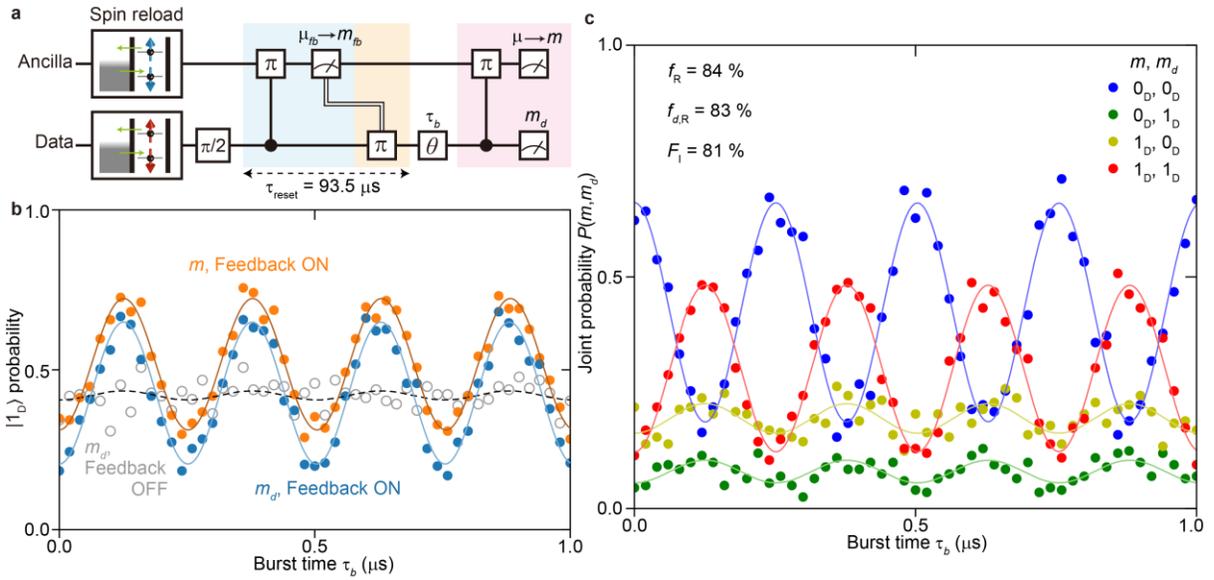

**Fig. 2 Test of the active reset. a,** Quantum circuit to test the reset protocol. The data and ancilla qubits are initialized at the beginning by reloading electrons from reservoirs. The data qubit is also subjected to a $\pi/2$ rotation before the active reset protocol. After the active reset, a resonant MW burst for $\tau_b$ is applied to the data qubit. Finally, the data qubit state is read out by a QND measurement and a destructive measurement (red area). **b,** Rabi oscillations measured with and without the feedback (solid and open circles, respectively) as a function of $\tau_b$. The solid and dashed curves are the fit curves for the data measured with and without the feedback, respectively. **c,** Joint probabilities for all four possible combinations of $m$ and $m_d$: $(m, m_d) = (0_D, 0_D)$, $(0_D, 1_D)$, $(1_D, 0_D)$, and $(1_D, 1_D)$ (blue, green, yellow, and red circles, respectively). The solid lines show fit curves to extract the initialization and readout fidelities.

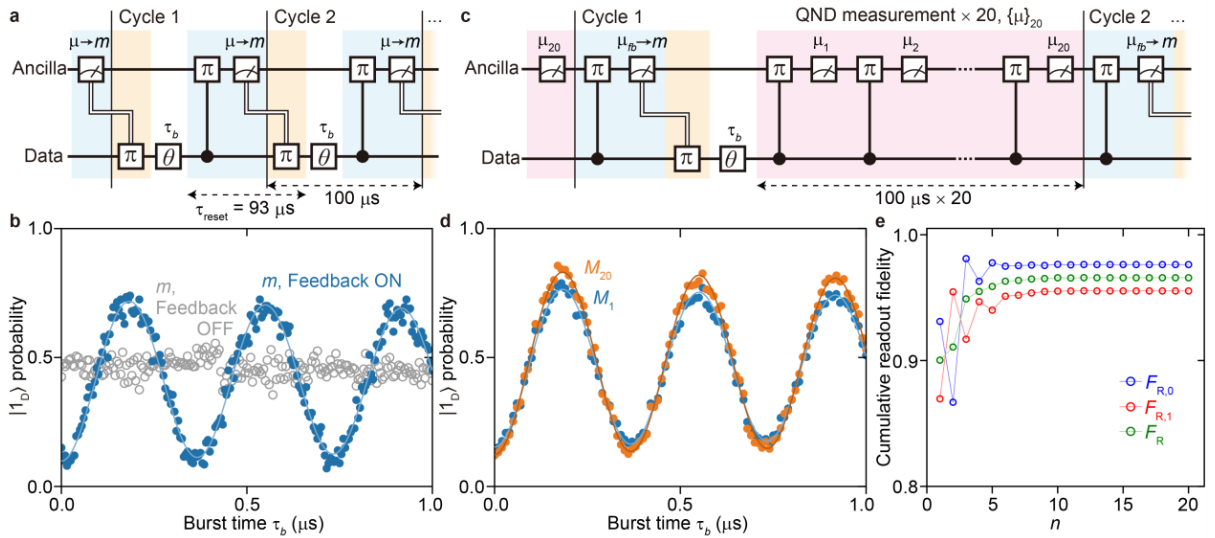

**Fig. 3 Active reset without electron reload to the data qubit and improvement of the visibility. a,** Quantum circuit to test the reset protocol without access of the data-qubit to the electron reservoirs. The ancilla qubit is initialized after each destructive measurement using access to the reservoir. **b,** Rabi oscillations measured with and without the feedback (solid and open circles, respectively). **c,** Quantum circuit with a cumulative readout sequence (red area) to measure the final data-qubit state. **d,** Rabi oscillations measured by the cumulative readout. Each plot is obtained by the Bayesian estimation using subsets $\{\mu\}_n$ ($n = 1$ and 20) of the set of repetitive measurement outcomes $\{\mu\}_{20}$ (blue and orange, respectively). The solid curves are sinusoidal fit. **e,** Cumulative readout fidelities as a function of $n$. The solid lines are eye guides.

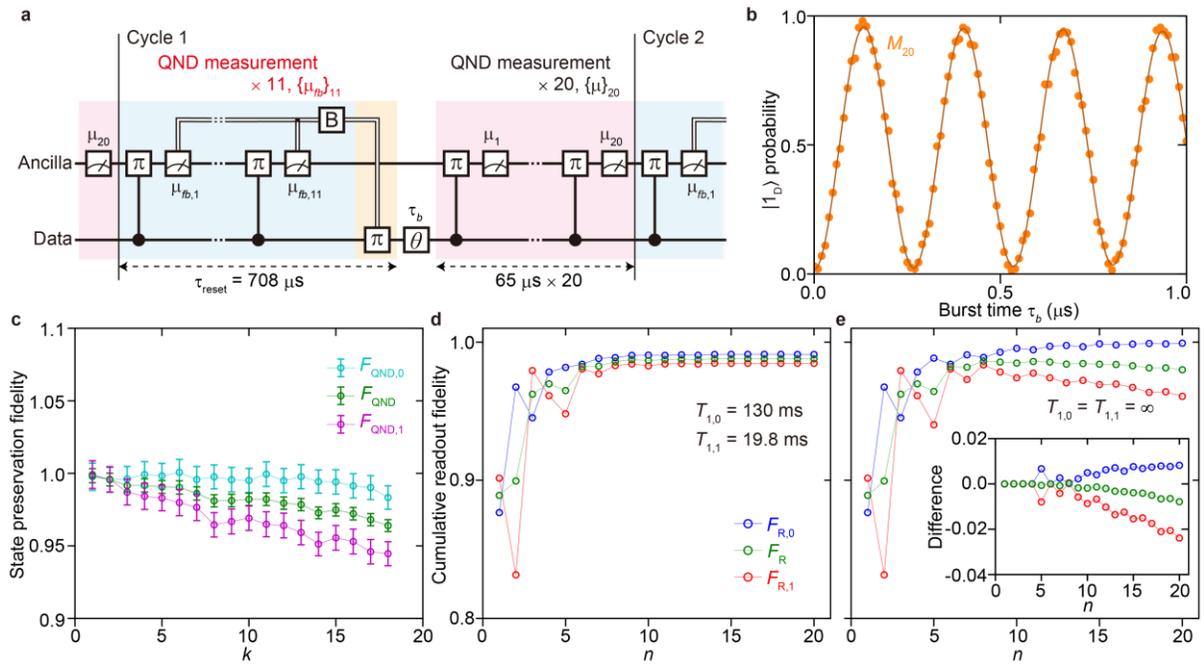

**Fig. 4 Feedback using a Bayesian-estimation logic. a,** Quantum circuit to test the reset protocol using a cumulative QND readout. Outcomes of the first 11 QND measurements (blue area) are fed to the Bayesian-estimation block (Box denoted B) and used to generate feedback. After the data qubit is reset and rotated resonantly, another cumulative readout sequence with 20 QND measurements (red area) is carried out to measure the data-qubit state. **b,** Rabi oscillations estimated by the Bayesian method using the set of ancilla readout outcomes $\{\mu\}_{20}$. The Bayesian estimation takes $(T_{1,0}, T_{1,1}) = (130\text{ ms}, 19.8\text{ ms})$ into account. **c,** State-preservation fidelities after a $k$-fold repetition of the QND measurements. The solid lines are eye guides. **d,e,** Cumulative readout fidelities estimated with assuming $(T_{1,0}, T_{1,1}) = (130\text{ ms}, 19.8\text{ ms})$ (d) and $(T_{1,0}, T_{1,1}) = (\infty, \infty)$ (e). The solid lines are eye guides. Inset: Difference between cumulative readout fidelities presented in d and e.